\documentclass{aa}

\usepackage{graphics}

\def\gt{{\rm \;>\;}}

\def\simgt{\lower.5ex\hbox{$\; \buildrel > \over \sim \;$}}

\begin{document}

   \thesaurus{04     
              (11.07.1 
               11.11.1 
               11.16.1)} 
   \title{On the axis ratio of the stellar velocity ellipsoid in disks 
of spiral galaxies}

   \author{P.C. van der Kruit
          \inst{1}
          \and
          R. de Grijs
          \inst{1,}
          \inst{2}
          }

   \offprints{P.C. van der Kruit}

   \institute{Kapteyn Astronomical Institute,
              University of Groningen,
              P.O. Box 800,
              9700 AV Groningen,
              the Netherlands,\\
              email: vdkruit@astro.rug.nl
         \and
             Astronomy Department,
             University of Virginia,
             P.O. Box 3818,
             Charlottesville, VA 22903,
             U.S.A.,\\
             email: grijs@virginia.edu
             }

   \date{Received 4 May 1999; accepted 29 September 1999}

   \maketitle

   \begin{abstract}

The spatial distribution of stars in a disk of a galaxy can be described
by a radial scale length and a vertical scale height.  The ratio of these
two scale parameters contains information on the axis ratio of the
velocity ellipsoid, i.e.  the ratio of the vertical to radial stellar
velocity dispersions of the stars, at least at some fiducial distance
from the center.  The radial velocity dispersion correlates well with
the amplitude of the rotation curve and with the disk integrated
magnitude, as was found by Bottema (1993).  These relations can
be understood as the result of the stellar disk being (marginally)
stable against local instabilities at all length scales.  This is
expressed by Toomre's well-known criterion, which relates the sheer in
the rotation to a minimum value that the radial stellar velocity
dispersion should have for stability for a given surface density.  Via
the Tully-Fisher (1977) relation, the velocity dispersion then becomes related
to the integrated magnitude and hence to the scale length.  The vertical
velocity dispersion relates directly to the scale height through
hydrostatic equilibrium.  It can be shown that the ratio of the two
length scales relates to the axis ratio of the velocity ellipsoid only
through the Toomre parameter $Q$ and in particular does not require a
choice of the mass-to-light ratio or a distance scale.  \\
We have applied this to the statistically complete sample of edge-on
galaxies, for which de Grijs (1997) has performed surface photometry and has
determined the length scales in the stellar light distribution.  

   \end{abstract}

%

\section{Introduction}

It has been known for many decades that the distribution of the stellar
velocities in the solar neighbourhood is far from isotropic.  
A longstanding problem in stellar dynamics (or in recent times more
appropriately called galactic dynamics) has been the question of the
shape and orientation of the velocity ellipsoid (i.e.  the
three-dimensional distribution of velocities) of stars in disks of
spiral galaxies.  The local ellipsoid has generated debate and
research for at least a century.  There still is no
consensus on the question of the orientation of the longest axis of the
velocity ellipsoid (the ``tilt'' away from parallel to the plane) at
small and moderate distances from the symmetry plane of the Galaxy (but
see Cuddeford \& Amendt 1991a,b), which is of vital importance for
attempts to estimate the local surface density of the Galactic plane
from vertical dynamics of stars.  The radial versus tangential dispersion
ratio is reasonably well understood as a result of the local sheer,
which through the Oort constants governs the shape of the epicyclic
stellar orbits (but see Cuddeford \&\ Binney 1994 and Kuijken \&\
Tremaine 1994 for higher order effects as a result of deviations from
circular symmetry).

There are two general classes of models for the origin of the velocity
dispersions of stars in galactic disks.  The first, going back to
Spitzer \& Schwarzschild (1951), is scattering by irregularities in the
gravitational field, later identified with the effects of Giant
Molecular Clouds (GMCs).  The second class of models can be traced back
to the work of Barbanis \& Woltjer (1967), who suggested transient
spiral waves as the scattering agent; this model has been extended by
Carlberg \& Sellwood (1985). Recently, 
the possiblity of infall of satellite galaxies has been recognized as a third 
option (e.g. Vel\'azquez \&\  White, 1999).

In the solar neighbourhood the ratio of the radial and vertical velocity
dispersion of the stars $\sigma _{\rm z}/\sigma _{\rm R}$ is usually
taken as roughly 0.5 to 0.6 (Wielen 1977;
see also Gomez et al.  1990), although values
on the order of 0.7 are also found in the literature (Woolley et al. 
1977; Meusinger et al.  1991).  The value of this ratio 
can be used to test predictions for the
secular evolution in disks and perhaps distinguish between the 
general classes of models.  Lacey (1984) and Villumsen (1985) have
concluded that the Spitzer-Schwarzschild mechanism is not in agreement
with observations: the predicted time dependence of the 
velocity dispersion of a group of stars 
as a function of age disagrees with the observed age
-- velocity dispersion relation (see also Wielen 1977), while it would
not be possible for the axis ratio of the velocity ellipsoid $\sigma
_{\rm z}/\sigma _{\rm R}$ to be less than about 0.7 (but see
Ida et al. 1993) 

Jenkins \& Binney (1990) argued that it is likely that the dynamical
evolution in the directions in the plane and that perpendicular to it
could have proceeded with both mechanisms contributing, but in different
manners.  Scattering by GMCs would then be responsible for the vertical
velocity dispersion, while scattering from spiral irregularities would
produce the velocity dispersions in the plane.  The latter would be the
prime source of the secular evolution with the scattering by molecular
clouds being a mechanism by which some of the energy in random motions
in the plane is converted into vertical random motions, hence
determining the thickness of galactic disks.  
The effects of a possible slow, but significant accretion of
gas onto the disks over their lifetime has been studied by
Jenkins (1992), who pointed out strong effects on the time dependence of
the vertical velocity dispersions, in particular giving rise to enhanced
velocities for the old stars. 

The only other galaxy in which a direct measurement of the velocity
ellipsoid has been reported, is NGC488 (Gerssen et al.  1997).  NGC488
is a moderately inclined galaxy, which enables these authors to solve
for the dispersions from a comparison of measurements along the major
and minor axes.  NGC488 is a giant Sb galaxy with a photometric
scale length of about 6 kpc (in the {\it B}-band) and an amplitude of the
rotation curve of about 330 km s$^{-1}$.  The axis ratio $\sigma _{\rm
z}/\sigma _{\rm R}$ is 0.70 $\pm $ 0.19; this value, which is probably
larger than that for the Galaxy, suggests that the spiral irregularities
mechanism should be relatively less important, in agreement with the
optical morphology. 

The light distribution in galactic disks
has in the radial direction an exponential behaviour
(Freeman 1970), characterised by a scale length $h$.  In the vertical
direction --at least away from the central layer of young stars and dust
that is obvious in edge-on galaxies-- it turns out that the light
distribution can also be characterised by an exponential scale height
$h_{\rm z}$, which is independent of galactocentric distance (van der
Kruit \& Searle 1981, but see de Grijs \& Peletier 1997).  It then is
usually assumed that the three-dimensional light distribution traces the
distribution of mass; this seems justifiable since the light measured is
that of the old disk population, which contains most of the stellar disk
mass and dominates the light away from the plane.  On general
grounds, these two typical length scales are expected to be independent,
the radial one resulting from the distribution of angular momentum in
the protogalaxy (e.g.  van der Kruit 1987; Dalcanton et al.  1997) or
that resulting from the merging of pre-galactic units in the galaxy's
early stages, while the length scale in the $z$-direction would result
from the subsequent, and much slower, disk heating and the consequent
thickening of the disk.  It is not {\it a priori} clear, therefore, that
the two scale lengths should correlate.  Yet, they do bear a relation to
the ratio of the velocity dispersions of the stars in the old disk
population.  The vertical one follows directly from hydrostatic
equilibrium.  In the radial direction it is somewhat 
indirect; a relation between the radial scale length and the
corresponding velocity dispersion comes about through conditions of
local stability (e.g.  Bottema 1993). 

In a recent study, de Grijs (1997, 1998; see also de Grijs \& van der
Kruit 1996) has determined the two scale parameters in a statistically
complete sample of edge-on galaxies and found the ratio of the two
($h$/$h_{\rm z}$) to increase with later morphological type.  In this
paper we will examine this dataset in detail in order to investigate
whether it can be used to derive information on the axis ratios of the
velocity ellipsoid and help make progress in resolving the general
issues described above. 

\section{Background}

In an extensive study of stellar kinematics of spiral galaxies, Bottema
(1993) presented measurements of the stellar velocity dispersions in the
disks of twelve spiral galaxies.  This first reasonably sized sample
represented a fair range of morphological types and luminosities,
although it was not a complete sample in a statistical sense.  In each
galaxy he determined a fiducial value for the velocity dispersion,
namely at one photometric ({\it B}-band) scale length.  He then found
that this fiducial velocity dispersion correlated well with the absolute
disk luminosity as well as with the maximum rotation velocity of the
galaxy.  His sample contained both highly inclined galaxies (where the
velocities in the plane are in the line of sight) and close to face-on
systems (where one measures the vertical velocity dispersion); when he
forced the relations for the two classes of galaxies to coincide he
found that a similar ratio between radial and vertical velocity
dispersion as applicable for the solar neighbourhood was needed. 

Bottema's empirical relation for velocity dispersion versus rotation
velocity is
\begin{equation}
\sigma _{\rm R,h} = 0.29\  V_{\rm rot},
\end{equation}
whereas for velocity dispersion versus disk luminosity it reads (in the
form of absolute magnitude)
\begin{equation}
\sigma _{\rm R,h} ({\rm km\ s}^{-1}) = -17 \times M_{B} - 279.
\end{equation} 

These relations can --for any galaxy for which the photometry is available
or for which the rotation curve is known-- be used to estimate the
radial stellar velocity dispersion of the old disk stars at one
photometric scale length from the center.  Doing this for the de Grijs
sample of edge-on galaxies and estimating the vertical velocity
dispersion from the vertical scale height, one can in principle determine
the axis ratio of the velocity ellipsoid for this entire sample.  It
would appear that this is a rather uncertain procedure, since one will
have to assume a mass-to-light ratio ($M/L$) in order to calculate the
vertical velocity dispersion from the photometric parameters.  We will
show, however, that $M/L$ 
does not enter explicitely in the formula for the ratio of velocity
dispersions. 

We will list our assumptions:\\
$\bullet $ The surface density of the disk has an exponential form as a
function of galactocentric distance:
\begin{equation}
\Sigma (R) = \Sigma (0)\ {\rm e}^{-R/h}.
\end{equation}
$\bullet $ The vertical distribution of density can be approximated by that
of the isothermal sheet (van der Kruit \& Searle 1981), but we will use
instead the subsequently suggested modification (van der Kruit 1988)
\begin{equation}
\rho (R,z) = \rho (R,0)\ {\rm sech} (z / h_{\rm z}).
\end{equation}
A detailed investigation of the sample (de Grijs et al.  1997) shows
indeed that the vertical light profiles are much closer to exponential
than to the isothermal solution, although the mass density distribution
most likely is less peaked than that of the light, since young
populations with low velocity dispersions add significantly to the
luminosity but little to the mass. 
Then the vertical velocity dispersion $\sigma _{\rm z}$ can be
calculated from
\begin{equation}
\sigma ^{2}_{\rm z} = 1.7051 \pi G \Sigma (R) h_{\rm z}.
\end{equation}
The usual parameter $z_{\circ}$ used in the notation for the isothermal
disk (and in de Grijs 1998) is $z_{\circ}=2h_{\rm z}$.  It is important
to note that this formula assumes that the old stellar disk is
self-gravitating.  Although this can be made acceptable for galaxies
like our own at positions of a few radial scale lengths from the center
(see van der Kruit \& Searle 1981), it is improbable in late-type
galaxies, which have significant amounts of gas in the disks, and we will
need to allow for this.\\ 
$\bullet $ The mass-to-light ratio $M/L$ is constant as a function of radius. 
Support for this comes from the observation by van der Kruit \& Freeman
(1986) and Bottema (1993) that the vertical velocity dispersion in
face-on spiral galaxies falls off with a scale length about twice that of
the surface brightness (but note that Gerssen et al.  1997, could not
confirm this for NGC488), combined with the observed
constant thickness of disks with galactocentric radius.\\
$\bullet $ We are not making any assumptions on the functional form of the
dependence of the radial velocity dispersion or the axis ratio of the
velocity ellipsoid. The observed radial stellar velocity dispersions 
in Bottema's sample are consistent
with a drop-off $\propto {\rm exp}(-R/2h)$, in which case
this axis ratio would be constant with galactocentric distance. 
However, over the range considered the data can be
fitted also with a radial dependence for the radial velocity dispersion
in which the parameter $Q$ for local stability against axisymmetric
modes (Toomre 1964) is constant with radius ($\propto R\ {\rm exp} (-R/h)$;
see van der Kruit \&
Freeman 1986).  The definition of Toomre's (1964) parameter $Q$ for
local stability against axisymmetric modes is
\begin{equation}
Q = {{\sigma_{\rm R} \kappa } \over { 3.36 G \Sigma}}.
\end{equation}
Disks are stabilised at small scales through the Jeans criterion by
random motions (up to the radius of the Jeans mass) and for larger
scales by differential rotation.  Toomre's condition states that the
minimum scale for stability by differential rotation should be no larger
than the Jeans radius.\\
$\bullet $ We assume that spiral galaxies have flat rotation curves with an
amplitude $V_{\rm rot}$ over all but their very central extent.  
This assumption implies that we may write the epicyclic frequency $\kappa$ as
\begin{equation}
\kappa = 2 \sqrt{B(B-A)} = \sqrt{2} {V_{\rm rot} \over R},
\end{equation}
where $A$ and $B$ are the Oort constants.\\

First we
will look into the background of the Bottema relations (1) and (2)
(see also van der Kruit 1990; Bottema 1993, 1997). 

Evaluating Toomre's $Q$ at $R= 1h$ and using the expression for the
epicyclic frequency above, we find
\begin{equation} 
\sigma_{\rm R,h} = { {3.36\ G} \over \sqrt{2}} Q {{\Sigma (h) h} \over
  V_{\rm rot}}.
\end{equation} 
Using $\Sigma (0) = (M/L) \mu _{\circ}$ and the total disk luminosity from
$L_{\rm d} = 2 \pi  \mu _{\circ } h^{2}$ we get
\begin{equation}
\sigma _{\rm R,h} = { {1.68 G} \over {{\rm e} \sqrt{\pi }}} Q \left( {M \over L} 
\right) {{\mu ^{1/2}_{\circ} L^{1/2}_{\rm d}} \over V_{\rm rot} }.
\end{equation}

Neither in the sample of galaxies that Bottema used to define his
relations, nor in our sample of edge-on systems do we have galaxies with
unusually low surface brightness.  It seems therefore justified to
assume that for the galaxies considered we have a reasonably constant
central surface brightness (Freeman 1970; van der Kruit 1987)
\begin{equation}
\mu _{\circ} \approx 21.6\ \  \mu_{B} = 142\  
{\rm L}_{\odot}\  {\rm pc}^{-2},
\end{equation}
where $\mu_{B}$ stands for {\it B}-magnitudes arcsec$^{-2}$. So, if $\mu 
_{\circ}$, $Q$ and $(M/L)$ are constant between galaxies, we see that the
fiducial velocity dispersion depends only on the disk luminosity and the
rotation velocity.

Bottema's relation (1) can then be reconciled with Eq. (9), if we have  
\begin{equation}
L_{\rm d} \propto  V^{4}_{\rm rot}.
\end{equation}
This is approximately the Tully-Fisher relation (Tully \& Fisher 1977);
not precisely, since we use the disk luminosity and not that of the
galaxy as a whole (however, for late-type galaxies this would be a minor
difference)\footnote{Eqs. (1) and (2) 
together would imply an exponential Tully-Fisher
relation rather than a power law. The problem is that Bottema chose to
fit a linear relation to his data, which is completely justified in view
of his error bars. But eqs. (11) and (9) would imply that he should have
fitted a curve in which the velocity dispersion is proportional to
$L_{\rm d}^{1/4}$ (see van der Kruit 1990, p. 199). Performing such a
fit to Bottema's data gives a curve that is only marginally
different from a straight line over his range of absolute magnitudes
(only a few km s$^{-1}$).
So, Eq. (2) should only be seen as an empirical fit of the data to a straight
line, although these data are equally consistent with the dependence
following from Eqs. (9) and (11).}.

So, we see that Bottema's relation (1) follows directly from Toomre's
stability criterion in exponential disks with flat rotation curves as
long as Eq. (11) holds. The proportionality constant in Eq. (11) can be
fixed using the parameters for the Milky Way Galaxy and for NGC 891 
as given in van der Kruit (1990).  These
two galaxies have $L_{\rm d} \sim 1.9\ \times \ 10^{10}\ {\rm
L}_{\odot}$ and $V_{\rm rot} \sim 220 \ {\rm km}\ 
{\rm s}^{-1}$\footnote{The distance
for NGC 891 --as are all distances in this paper-- 
is based on a Hubble constant of 75 km s$^{-1}$ Mpc$^{-1}$.}. This
gives
\begin{equation}
L_{\rm d}\ ({\rm L}_{\odot}) = 8.11\ V^{4}_{\rm rot}\ ({\rm km}\ {\rm
s}^{-1}). 
\end{equation}
and using also Eq. (10) we get
\begin{equation}
\sigma _{\rm R,h} = 5.08 \times 10^{-2} Q \left( {M \over L} \right)
V_{\rm rot}. 
\end{equation}
>From this we find with Bottema's relation (1), that 
$Q (M / L)_{B} \approx 5.7$. 
In a somewhat different, but comparable manner, Bottema (1993) has
also concluded that this product is of order 5.

\bigskip

We now turn to the vertical velocity dispersion. 
Evaluating the equation for hydrostatic equilibrium (5) at  
galactocentric distance $R=1h$ we find
\begin{equation}
\sigma _{\rm z,h} = \left\{ 5.36\ G \Sigma (h) h_{\rm z} \right\}^{1/2},
\end{equation}
and can thus calculate the vertical velocity dispersion from 
\begin{equation}
\sigma _{\rm z,h} = \left\{ {5.36 \over {\rm e}}\ G \left( {M \over L}
\right) \mu _{\circ } h_{\rm z} \right\}^{1/2}. 
\end{equation}
\bigskip

Finally we examine the ratio of the two velocity dispersions.
If we eliminate $\Sigma (h)$ between Eqs. (8) and (14) we obtain
\begin{equation}
\sigma _{\rm R,h} = 0.444\ Q { \sigma_{\rm z,h}^{2} \over V_{\rm rot}}
{h \over h_{\rm z}},
\end{equation}
and with Eq. (1) 
\begin{equation}
\left( {\sigma _{\rm z} \over \sigma _{\rm R}} \right)_{\rm h}^{2} =
{7.77 \over Q} 
{h_{\rm z} \over h}.
\end{equation}

Note that due to the elimination of the surface density also the
mass-to-light ratio has dropped out of this equation and the result is
independent of any assumption on $M/L$.
Eq.  (17) translates the ratio of the two length scales to that of the
corresponding velocity dispersions and the underlying physics can be
summarized as follows.  In the vertical direction the length scale and
the velocity dispersion relate through dynamical equilibrium.  In the
radial direction the velocity dispersion is related to the epicyclic
frequency through the local stability condition, which is proportional
to the rotation velocity.  The ``Tully-Fisher relation'' then relates
this to the integrated magnitude and hence to the size and length scale
of the disk. 

One should be careful in the use of Eq. (17), since in practice
photometric scale lengths are wavelength dependent and its derivation 
--and therefore the numerical constant-- is valid
only at one exponential {\it surface density} scale length. The purpose of
presenting it here is only to show that, if the two velocity dispersions are
derived in a consistent manner from Eqs. (8) and (14) --or alternatively
Eqs. (9) and (15)--, the assumption used for $M/L$ drops
out in the resulting ratio.

\section{Application to the de Grijs sample}

The sample of edge-on disk galaxies of de Grijs (1998) contains 46
systems for which the structural parameters of the disks have been
determined (including a bulge/disk separation in the analysis).  From
this sample we take those for which rotation velocities have been
derived in a uniform manner (Mathewson et al.  1992; data collected in
de Grijs 1998, Table 4) as well as those for which the Galactic
foreground extinction in the {\it B}-band is less than 0.25 magnitudes. 
For this remaining sample of 36 galaxies we perform the following
calculations:\\
$\bullet $  From the total magnitudes in de Grijs (1998, Table 6) we obtain
the integrated {\it B}- and {\it I}-magnitudes of the disk.\\
$\bullet $  Using the radial scale length as measured in the {\it I}-band we
calculate the central (face-on) surface brightness of the disk from its
{\it I}-band integrated luminosity. So, we do not use Eq. (10)
for a constant central surface brightness.\\
$\bullet $  Then we use one or both of the two Bottema relations (1) and (2)
to estimate the radial velocity dispersion at one photometric
scale length (by definition in the {\it B}-band). 
Where we can do this with both relations, the ratio between the two
estimates is 1.11 $\pm $ 0.19.  This is not trivial, as the rotation
velocities and disk luminosities are determined completely independently 
(and by different workers) and only the one using
the absolute magnitude needs an assumption for the
distance scale. \\
$\bullet $  Then using Eq.  (15), we estimate the vertical velocity dispersion
at one photometric scale length in the {\it I}-band. 
For this we need a value for the mass-to-light ratio and we will discuss
this first.
\bigskip

We found, that through Eq. (13) Bottema's relation (1) provides 
a value for $Q (M/L)$ of about 5.7. 
So we make a choice for $Q$ rather than for $M/L$.  
It has become customary to assume values of $Q$ of
order 2, mainly based on the numerical simulations of Sellwood \&
Carlberg (1984), who find their disks to settle with $Q \sim $ 1.7 at
all radii.  In principle we can use the observed properties of the
Galaxy to fix $Q$ from Eq.  (17).  We have $(\sigma _{\rm z}/\sigma
_{\rm R})^{2} \sim 0.5$ (in the solar neighbourhood, but assume for the
sake of the argument also at $R = 1h$) and $h_{\rm z}/h \sim 0.1$ (see
Sackett 1997 for a recent review), so that indeed $Q \sim 1.7$.  We
will make the general assumption that $Q$ = 2, in agreement with the
considerations above; then $(M/L)_{B}$ = 2.8. 

The rotation velocity version of Bottema's empirical relations (Eq. 
(1)) can provide further support for the choice of $Q$, along the lines
of the discussion in van der Kruit \& Freeman (1986).  In the first
place we recall the condition for the prevention of swing amplification
in disks (Toomre 1981), as reformulated by Sellwood (1983)
\begin{equation}
X = {{R \kappa ^{2} \over {2 \pi m G \Sigma }}} \gt 3,
\end{equation}
where $m$ is the number of spiral arms.
For a flat rotation curve this can be rewritten as
\begin{equation}
{ {Q V_{\rm rot} \over \sigma _{\rm R}}} \gt 3.97 m.
\end{equation}
With Eq. (1) this becomes $Q \gt 1.15 m$.
Considering that the coefficient in Eq.  (1) has an uncertainty of
order 15\%, this tells us that we have to assume $Q$ at least
of order 2 to prevent strong barlike ({\it m}=2) disturbances in the
disk. 

A similar argument can be made using the global stability criterion of 
Efstathiou et al. (1982). This criterion states that for a galaxy
with a flat rotation curve and an exponential disk, global stability 
requires a dark halo and
\begin{equation}
Y = V_{\rm rot} \left( {h \over {G M_{\rm disk}}} \right)^{1/2} \simgt 1.1.
\end{equation}
Here $M_{\rm disk}$ is the total mass of the disk. This can be rewritten
as
\begin{equation}
Y = 0.615 \left[ {{Q R V_{\rm rot}} \over {h \sigma _{\rm R}}} \right]^{1/2}
{\rm exp} \left( -{R \over {2h}} \right) \simgt 1.1,
\end{equation}
and, when evaluated at $R = 1h$, yields with Eq. (1) $0.69 \sqrt{Q} 
\simgt 1.1$,
and therefore also implies that $Q$ should be at least about 2. 
Efstathiou et al.  have also come to this conclusion for our Galaxy,
with the use of local parameters for the solar neighbourhood. 
\bigskip

Having adopted a value for $Q$ and through this a value for $(M/L)_{B}$,
we will have to convert it to $(M/L)_{I}$.  For this we need a $(B - I)$
colour for the disks.  From the fits of de Grijs (1998) we find that the
total {\it disk} magnitudes show a rather large variation in colour; for
the sample used here $(B - I)$ has a mean value of 1.9, but the r.m.s. 
scatter is 0.8 magnitudes.  In his discussion, de Grijs (1998) suspects
a systematic effect of the internal dust in the disks (particularly on
the {\it B}-magnitudes, which is another reason for us to use the
$V_{\rm rot}$-version of the Bottema relation (Eq. (1))
in our derivation in the
previous section).  Instead we turn to the discussion of de Jong
(1996b), who compares his surface photometry of less inclined spirals to
star formation models.  From his Table 3, we infer that for single burst
models with solar metallicity and ages of 12 Gyr $(M/L)_{B} = 2
(M/L)_{I}$.  So we will use an $(M/L)_{I}$ of 1.4. 
\bigskip

There is a further refinement required. In order to
take into account the fact that in late-type galaxies the gas
contributes significantly to the gravitational force, we have to correct
for a galaxy's gas content as a function of Hubble type. In the
following, we will discuss the observational data regarding the H{\sc i}
and the H$_2$ separately.

\begin{figure*}
\resizebox{\hsize}{!}{\includegraphics{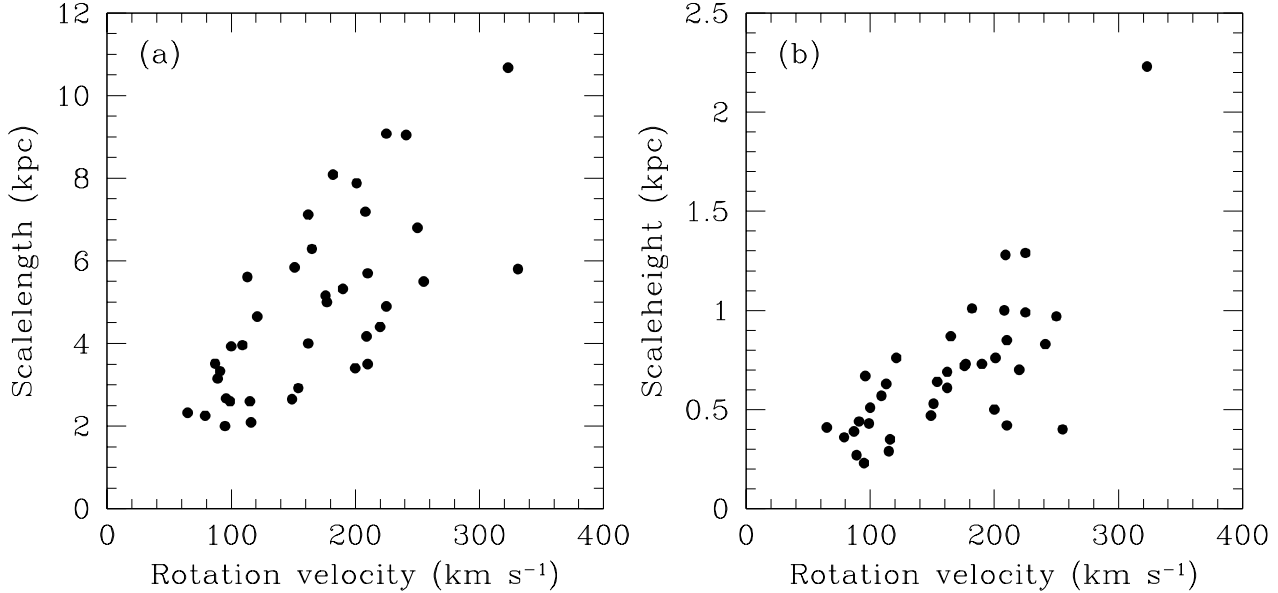}}
\vspace{-9.5cm}
\caption{The scale lengths (a) and scale heights (b) of the galaxies in
our sample as a function of their rotation velocities.  The r.m.s. 
errors are of order 10 km s$^{-1}$ in the rotation velocities (Mathewson
et al.  1992), 5\% in the radial scale length and 9\%\ in the vertical
scale height (de Grijs 1998).}
\end{figure*}

For 25 of de Grijs' sample galaxies H{\sc i} observations are available,
so that we can estimate the gas-to-total disk mass. For this we apply a
correction of a factor 4/3 to the H{\sc i} in order to take account of
helium and use de Grijs' (1998) {\it I}-band photometry and our adopted
$M/L_B$ ratio of 2.8 (see below) to estimate the total disk mass. As a
function of Hubble type we then find
\begin{center}
\begin{tabular}{ l l l }
\hline
\hline
Type & gas-to-total disk mass & n \\
\hline
Sb  & 0.31 $\pm $ 0.17 & 3 \\
Sbc & 0.36 $\pm $ 0.19 & 5 \\
Sc  & 0.53 $\pm $ 0.09 & 5 \\
Scd & 0.49 $\pm $ 0.16 & 9 \\
Sd  & 0.52 $\pm $ 0.10 & 3 \\
\hline
\hline
\end{tabular}
\end{center}
We find no dependence on rotation velocity:
\begin{center}
\begin{tabular}{ l l l }
\hline
\hline
$V_{\rm rot}$ (km s$^{-1}$)& gas-to-total disk mass & n \\
\hline
\ 80 -- 130 & 0.51 $\pm $ 0.09 & 10 \\
130 -- 180 & 0.45 $ \pm $ 0.17 & 8 \\
180 -- 230 & 0.51 $\pm $ 0.09 & 7 \\
\hline
\hline
\end{tabular}
\end{center}
So, the H{\sc i} mass is about half the stellar mass in disks of Sb's
and about similar to that 
in Sc's and Sd's. But there is no dependence on rotation
velocity. But this is not what we need; we should use surface densities
rather than disk masses. Now, the H{\sc i} is usually more extended than
the stars and has a shallower radial profile. So the ratios in the
tables above are definite upper limits.
In order to take into account the effect that in
late-type systems the gas contributes significantly to the gravitational
force we have ``added'' for types Scd and Sd a similar amount of gas as
in stars and half of that for Sc's.

The distribution of H$_2$ in spiral galaxies is a more complex matter;
it is often centrally peaked, although some Sb galaxies exhibit central
holes (for a recent review, see Kenney 1997). The molecular fraction of
the gas appears to be lower in low-mass and late-type galaxies, assuming
that the conversion factor from CO to molecular hydrogen is 
universal\footnote{There is even doubt concerning the constancy of this
conversion factor {\it within} our Galaxy (Sodroski et al.
1995).}. Since our sample galaxies are generally low-mass, later-type
systems, we believe that the corrections for molecular gas are small,
and therefore contribute little to the correction for the presence of gas.

We added (a) the galaxies from van der
Kruit \& Searle (1982) to the sample, (b) our Galaxy using the Lewis
\& Freeman (1989) velocity dispersion and the structural parameters in
van der Kruit (1990), and (c) the observational results for NGC488 from
Gerssen et al.  (1997). We leave the few early type (S0
and Sa) galaxies out of the discussion, because the component seperation in
the surface brightness distributions is troublesome and some of our assumptions
(in particular the self-gravitating nature of the disks) are probably
seriously wrong. 

In order to be able to trace the origin of our results, we first show in
Fig.  1 the radial and vertical scale lengths of the sample as a function
of the rotation velocity.  Both increase with $V_{\rm rot}$, which would
be expected intuitively.  The main result is presented in Figs.  2 and
3.  From Fig.  2 we see that the vertical velocity dispersions, that
have been derived from hydrostatic equilibrium, increase with the
rotation speed (the radial velocity dispersions do the same
automatically as a result of the use of the Bottema relations).  For the
slowest rotation speeds the predicted vertical velocity dispersion is on
the order of 10-20 km s$^{-1}$, which is close to that observed in the
neutral hydrogen in face-on galaxies (van der Kruit \& Shostak 1984). 

\begin{figure*}
\resizebox{\hsize}{!}{\includegraphics{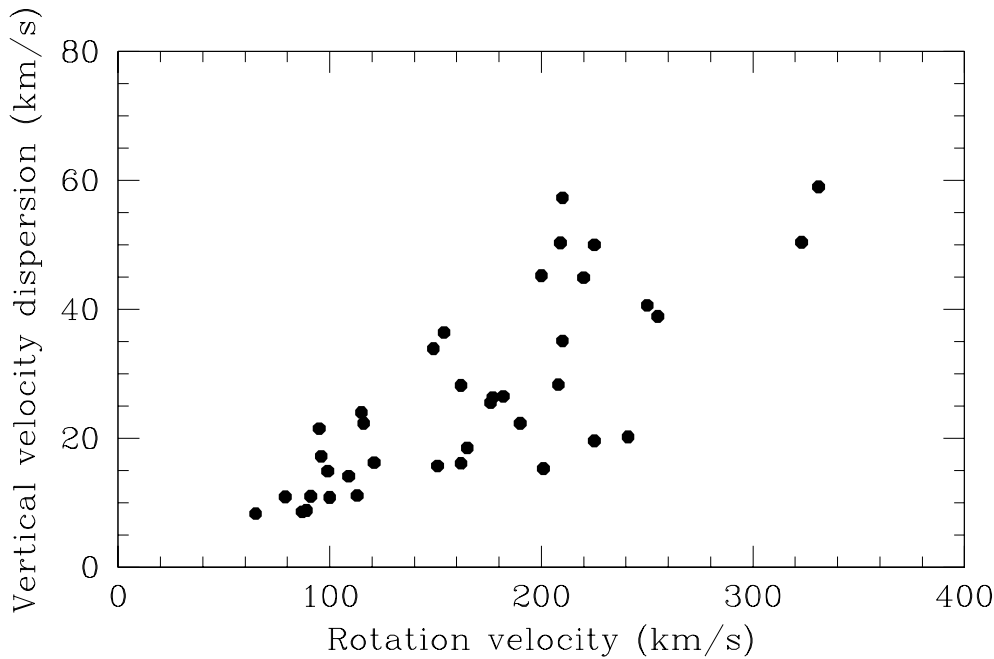}}
\vspace{-1cm}
\caption{The calculated vertical velocity dispersions at one exponential 
scale length from the center as a function of the rotation velocities}
\end{figure*}

The distribution of the axis ratio of the velocity ellipsoid
with morphological type is as follows:
\begin{center}
\begin{tabular}{ l l l }
\hline
\hline
Type & $\sigma _{\rm z,h} / \sigma_{\rm R,h}$ & n \\
\hline
Sb  & 0.71 $\pm $ 0.14 & 11 \\
Sbc & 0.69 $\pm $ 0.16 & 7 \\
Sc  & 0.49 $\pm $ 0.17 & 6 \\
Scd & 0.70 $\pm $ 0.20 & 11\\
Sd  & 0.63 $\pm $ 0.22 & 5 \\
\hline
\hline
\end{tabular}
\end{center}
Not much of a trend is seen here. 
It is in order to comment here briefly on the effects of our corrections
for the gas to obtain vertical velocity dispersions. From our discussion
above we conclude that there would be no
systematic effect introduced as a function of rotation velocity.
Furthermore, taking away our correction altogether
reduces the values for the
average axis ratio in the table just given to about 0.55 for Scd and Sd
galaxies. Even in this unrealistic case of not allowing for the presence
of the gas, we believe the trend to be hardly significant in view of the
uncertainties. Since some correction for the gas mass 
as a function of morphological type must
be made, we cannot claim that we find any evidence for a change in the velocity
anisotropy with Hubble type.

Fig. 3 shows the axis ratio of the velocity ellipsoid of all the galaxies
versus their rotation velocity.  In view of the
fact that the two dispersions that go into this ratio are determined
from different observational data ($\sigma _{\rm R,h}$ from integral
properties such as total luminosity and amplitude of the rotation curve;
$\sigma _{\rm z,h}$ from photometric scale parameters and surface
brightness) and that we have made rather simplifying assumptions, the
scatter is remarkably small.  No systematic trends are visible (and
would probably not be significant!) in the data.  The points closest to
unity in the dispersion ratio generally have low rotation velocities and
inferred velocity dispersions.  One of these ($V_{\rm rot}$ = 95 km
s$^{-1}$, $\sigma _{\rm z,h} / \sigma _{\rm R,h}$ = 0.75) is NGC5023. 
Bottema et al.  (1986) have shown that the stars and the gas in this
galaxy are effectively coexistent; the radial {\it and} vertical
distributions are very similar.  This would imply that the velocity
dispersions of the gas and the stars are the same.  The vertical velocity
dispersion found here is about 20 km s$^{-1}$, which is significantly
higher than that observed in larger spirals.  It would be of interest to
measure the H{\sc i} velocity dispersion in this galaxy.  Since the
HI would be expected to have an isotropic velocity distribution from
collisions between clouds, the vertical dispersion should be equal to that
in the line of sight in edge-on galaxies.

\begin{figure*}
\resizebox{\hsize}{!}{\includegraphics{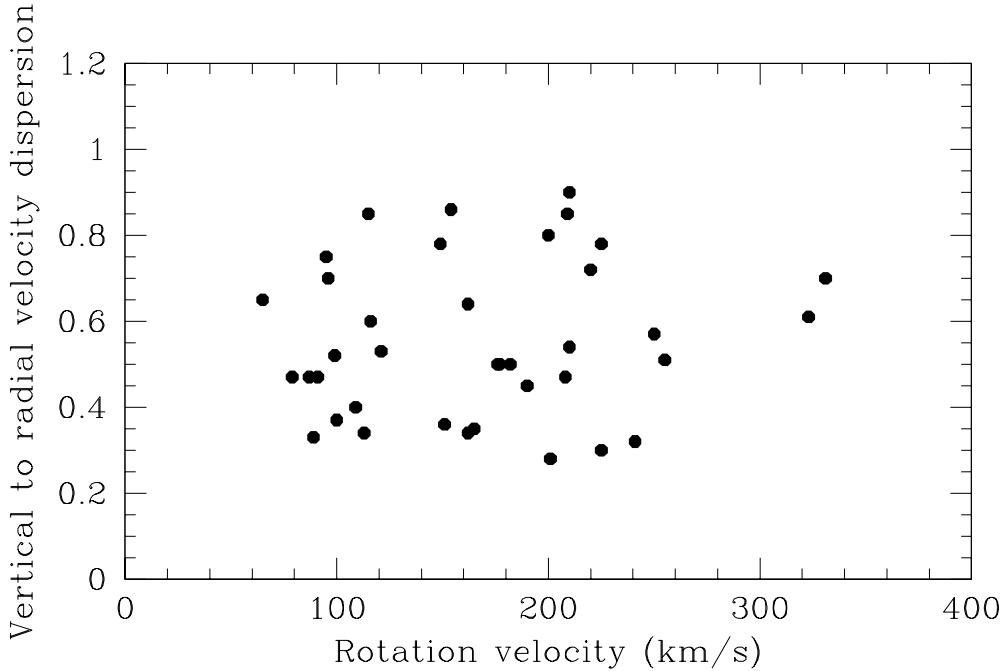}}
\vspace{-1cm}
\caption{The ratio of the calculated vertical to radial velocity
dispersions at one exponential scale length from the center as a function
of the rotation velocities}
\end{figure*}

\section{Discussion}

In this section we will critically discuss the uncertainties in our
approach.  

$\bullet $ {\it The linearity of the magnitude version of the Bottema
relation.} We have discussed above that the power-law nature of the
Tully-Fisher relation (Eq. (11)) would imply a nonlinear form of the
magnitude version of Bottema's relation (Eq. (2)). We 
have used it as an empirical
relation to help (together with Eq. (1)) to estimate the radial
velocity dispersions from the observed photometry. One may argue that it
is internally consistent to use instead of Eq. (2) a fit of the form 
\begin{equation}
\sigma _{\rm R,h} \propto  L_{\rm d}^{1/4}.
\end{equation} 
This has only a noticable effect on galaxies with faint absolute disk
magnitudes. We have repeated our analysis using such a fit and find no
change in our results. To be more definite we repeat the table of
the average axis ratio as a function of morphological type that we then
obtain.
\begin{center}
\begin{tabular}{ l l l }
\hline
\hline
Type & $\sigma _{\rm z,h} / \sigma_{\rm R,h}$ & n \\
\hline
Sb  & 0.67 $\pm $ 0.17 & 11 \\
Sbc & 0.64 $\pm $ 0.13 & 7 \\
Sc  & 0.47 $\pm $ 0.17 & 6 \\
Scd & 0.60 $\pm $ 0.11 & 11\\
Sd  & 0.60 $\pm $ 0.20 & 5 \\
\hline
\hline
\end{tabular}
\end{center}

$\bullet $  {\it Choice of $Q$}.  We have adopted a value for $Q$ of 2.0, and
this enters directly in our results in the calculation of the vertical
velocity dispersion through the value of $M/L$ that follows from this
choice.  Had we adopted a value of 1.0 for $Q$, $M/L$ would have been a
factor 2 higher and the value for $\sigma _{\rm z,h}$
a factor $\sqrt{2}$ --see Eq.  (15)--.  Bottema (1993, his Fig. 11)
has shown that his observations of the stellar kinematics 
do not show any evidence for systematic variations in $Q$ among
galaxies.\\
We have used Eq. (19) to argue that $Q$ is likely of order 2 in order to 
prevent strong barlike (m=2) disturbances. It does not 
necessarily follow from
this that galaxies with more spiral arms should have higher values of
$Q$ or that $Q$ should be significantly
lower in galaxies with very strong two-armed structure. Spiral structure 
may arise in a variety of ways; we only argue that disks do not have grossly 
distorted m=2 shapes and that therefore swing amplification is apparently 
not operating. \\
We have assumed that the stellar disk is self-gravitating and ignored the
influence of the gas in the evaluation of $Q$. At one scale length this
is probably justified (it is only a small effect in solar
neighbourhood), even for late-type disks. \\
$\bullet $  {\it Non-exponential nature of the disks}.  Often the disk in
actual galaxies can be fitted to an exponential only over a limited
radial extent.  In that case our description is unlikely to hold. 
However, in our sample the fits can be made reasonably well at one
scale length from the center and we believe this not to be a problem. \\
$\bullet $  {\it Vertical structure of the disks}.  Our results depend
on the adoption of a particular form of the vertical mass distribution
(namely the sech($z$)-form) of Eq.  (4).  This enters our results
through the value of the numerical constant 1.7051 in Eq.  (5), and in
Eq.  (15) it enters into the value for $\sigma _{\rm z,h}$ as its square
root.  Had we assumed the isothermal distribution, then the constant
would have been 2.0, while it would have been 1.5 for the exponential
distribution.  This would have given us values for $\sigma _{\rm
z,h}$ which are only 6 to 8\%\ higher or lower.  
As we have shown in de Grijs et
al. (1997), for our sample galaxies the vertical {\it luminosity}
distributions in these disk-dominated galaxies are slightly rounder than
or consistent with the exponential model. However, the vertical {\it
mass} distribution is probably less sharply peaked, and thus expected to
be more closely approximated by the sech({\it z}) model.\\
$\bullet $  {\it Non-constancy of central (face-on) surface brightness}.  We
have assumed in section 2 that for all galaxies the central surface
brightness is constant.  This is certainly unjustified for
so-called ``low surface brightness galaxies''; however, our galaxies
have brighter surface brightnesses than galaxies that are usually
considered to be of this class.  But even for galaxies as in our sample
it remains true that the (face-on) central surface brightness is in
general somewhat lower for smaller systems (van der Kruit 1987; de Jong
1996a).  From de Jong's bivariate distribution functions, it can be seen
that late-type, low absolute mag\-nitude spirals may have a central
surface brightness (in {\it B}) that is up to 1.0 magnitude fainter. 
Note, however, that for our derived velocity dispersons 
we have used the {\it actually
observed} surface brightness for each galaxy. \\
$\bullet $  {\it Effects of colour variations on the mass-to-light ratio}. 
There is a fairly large variation in the colours of the disks in the
sample.  De Grijs (1998) has argued that this is the result of internal
dust extinction.  However, we have used the {\it I}-band data; de Grijs'
Fig.  11 shows that the variation is much less in $(I - K)$ than in
colours involving the {\it B}-band.\\
The colours of the disks do not correlate with morphological type (de
Grijs 1998); although such a correlation has been seen in more face-on
galaxies (de Jong 1996b), we believe that the fitting procedure has
ignored most of the young population and dust absorption near the plane,
and that contributions to the $M/L$ scatter as a result of young
populations has mostly been avoided; and that the {\it I}-band
scale lengths determined away from the galactic planes are fairly
representative of the stellar mass distributions (de Grijs 1998).\\ 
The colour of the
fitted disks in our sample has $(I - K)$ in the range $\sim $ 2 to 4. 
The latter is red, even for an old population and may well be caused by
excessive internal extinction, but we see no strong evidence for a
substantial systematic correction in our velocity dispersions from this. \\
$\bullet $  {\it Effects of metallicity on the mass-to-light ratio}.  De Jong
(1996b) has drawn attention to the non-negligible effects of metallicity
on the mass-to-light ratio.  From his compilation of models, in
particular his W94 (Worthey 1994) models with ages of 12 Gyr, we
estimate that the effect in the {\it I}-band amounts to 10 to 20\%\ over
the range of relevant metallicities.  The effect on the derived velocity
dispersions is the square root of this. \\
$\bullet $  {\it Effects of radial colour variations}.  Since we use an
empirical relation to derive the radial velocity dispersion at one {\it
B}-scale length from the center, we have to consider the effect of using
the {\it I}-band.  We have used the latter as the proper scale length to
use for the mass density distribution.  The correct one to use here
would be the one measured in the {\it K}-band, which is 1.15 $\pm $ 0.19
times smaller for this sample.  The scale length in the {\it B}-band is
1.64 $\pm $ 0.41 times longer than in the {\it K}-band (values quoted
here from de Grijs 1998).  We deduce from this that we may have
underestimated the scale length to use by a factor 1.43 and systematically
overestimated the vertical velocity dispersion by about 20\%.\\
The effect of radial metallicity variations on the scale length is
probably very small; these gradients in the older stellar populations
are in any case expected to be significantly less than in the
interstellair medium. This is so, because in models for galactic chemical
evolution the mean stellar metallicity of the stars approximates the
(effective) yield, while that in the gas grows to much large values in
most models (van der Kruit 1990, p. 322).\\
$\bullet $  {\it Non-flatness of rotation curves}.  This may be an effect for
small, late-type galaxies that have slowly rising rotation curves. 
It enters however in our analysis only in the derivation of the Bottema
relations and this holds empirically to rather small rotation velocities
($\sim $ 100 km s$^{-1}$). \\
$\bullet $  {\it The slope of the Tully-Fisher relation}.  Although we use a
relation between the luminosity of the disk alone and the rotation
velocity, it remains true that our ``slope'' of $V_{\rm rot}^4$ is
steeper than the usually derived slopes of Tully-Fisher relations, which
would indicate $V_{\rm rot}^3$ (e.g.  Giovanelli et al.'s 1997
``template relation'' yields an exponent of 3.07 $\pm $ 0.05).  If this
slope were to be put into Eq.  (9), Eq.  (1) would have $\sigma _{\rm
R,h} \propto \sqrt{V_{\rm rot}}$, which is very significantly in
disagreement with Bottema's observations.  The same holds for the
other Bottema relation, Eq.  (2).\\ 
The analysis of
the present sample (see de Grijs \&\ Peletier 1999) has resulted in slopes in
the Tully-Fisher relation of 3.20 $\pm $ 0.07 in the {\it I}-band and
3.24 $\pm $ 0.21 in the {\it K}-band.  On the other hand, Verheijen
(1997), in his extensive study of about 40 galaxies in the Ursa Major
cluster, finds a slope of 4.1 $\pm $ 0.2 in the {\it K}-band. \\
$\bullet $  {\it Effects of the gas on the value of $Q$}.  In our calculations
we have already made crude allowance for the effects of the gas on the
gravitational field.  But in our derivations we have not taken into
account the effect of the H{\sc i} on the effective velocity dispersion to be
used in the evaluation of the $Q$-parameter.  The effect of the H{\sc i} is a
decrease of the effective velocity dispersion and therefore in $Q$. 
This means that the assumed value should in reality be decreased
on average, but beyond this numerical effect, it does not affect our
results. 

The effects just discussed can produce errors in the estimated velocity
dispersions of the order of 10 to 20\%\ each.  The final result of the
dispersion ratios in Fig.  3 may therefore be wrong by a few tenths,
which is comparable to the scatter in that figure.  However, we have no
cause to suspect that we have introduced serious systematic effects that
would be strong functions of the rotation velocity or the morphological
type and the lack of correlation of the axis ratio with these properties
is unlikely to be an artifact of our analysis.

We conclude that it is in principle possible to
infer information on the axis ratio of the velocity ellipsoid from a
sample of edge-on galaxies for which both the radial scale length as well
as the vertical scale height have been measured. The result, however
shows much scatter, most of which is a result of the necessary
assumptions. There is one
significant improvement that can be made and that is the direct
observation of the stellar velocity dispersion in these disks.  That
this is feasible in practice for edge-on systems has been shown by
Bottema et al.  (1987, 1991).  The observed velocity profiles can be
corrected for the line-of-sight effects, giving the tangential velocity
dispersion, which through the observed shape of the rotation curve can
be turned into the radial velocity dispersion.  Although a time
consuming programme, we believe that it is worth doing for two reasons:
(1) It will set both versions of the Bottema relation on a firmer
footing.  (2) The uncertainties in the analysis above can likely be
significantly diminished by direct measurement of the radial velocity
dispersion rather than having to infer it from the rotation velocity or
the disk absolute magnitude.

\section*{Acknowledgements}
PCvdK thanks Jeremy Mould for hospitality at the Mount Stromlo and
Siding Spring Observatories, where most of this work
was done, and Ken Freeman for discussions, the Space Telescope Science
Institute and Ron Allen for hospitality, when
the final version of this paper was prepared, and the Faculty of
Mathematics and Natural Sciences of the University of Groningen for
financial support that made these visits possible.
RdG was supported by NASA grants NAG 5-3428 and NAG 5-6403.

\end{document}